\documentclass[12pt,eqsecnum,nofootinbib,floats,preprint,aps,prd,floatfix,titlepage,superscriptaddress,tightenlines]{revtex4} 
\usepackage{epsfig}
\usepackage{amsmath}
\usepackage{amssymb}
\usepackage{graphics}
\usepackage{bm}

\begin{document}

\title{Generalized surface tension bounds in vacuum decay}
\author{Ali Masoumi}
\email{ali@cosmos.phy.tufts.edu}
\affiliation{Institute of Cosmology, Department of Physics and Astronomy,
Tufts University, Medford, MA 02155, USA}
\author{Sonia Paban}
\email{paban@physics.utexas.edu}
\affiliation{Department of Physics,
The University of Texas at Austin,\\  Austin, Texas 78712, USA}
\author{Erick J. Weinberg}
\email{ejw@phys.columbia.edu}
\affiliation{Physics Department, Columbia University, New York, New York 10027, USA}

\begin{abstract}

Coleman and De Luccia (CDL) showed that gravitational effects can
prevent the decay by bubble nucleation of a Minkowski or AdS false
vacuum.  In their thin-wall approximation this happens whenever the
surface tension in the bubble wall exceeds an upper bound proportional
to the difference of the square roots of the true and false vacuum
energy densities.  Recently it was shown that there is another type of
thin-wall regime that differs from that of CDL in that the radius of
curvature grows substantially as one moves through the wall.  Not only
does the CDL derivation of the bound fail in this case, but also its
very formulation becomes ambiguous because the surface tension is not
well-defined.  We propose a definition of the surface tension and show
that it obeys a bound similar in form to that of the CDL case.  We
then show that both thin-wall bounds are special cases of a more
general bound that is satisfied for all bounce solutions with
Minkowski or AdS false vacua.  We discuss the limit where the
parameters of the theory attain critical values and the bound is
saturated. The bounce solution then disappears and a static planar
domain wall solution appears in its stead.  The scalar field potential
then is of the form expected in supergravity, but this is only
guaranteed along the trajectory in field space traced out by the
bounce.
\end{abstract}

\preprint{UTTG-12-17}
\maketitle

\section{Introduction}

In their classic study of vacuum decay via bubble nucleation, Coleman
and De Luccia (CDL)~\cite{Coleman:1980aw} discovered a surprising
feature of decays from a Minkowski vacuum to an anti-de-Sitter (AdS)
vacuum or from one AdS vacuum to another.  If a potential has two
vacua of differing energy, decay from the higher energy false vacuum
to the lower energy true vacuum is always possible, if gravitational
effects are ignored.  However, if the higher vacuum has either zero or
negative energy, such decays are quenched if the two vacua are
sufficiently close in energy.  In the thin-wall approximation of CDL,
bubble nucleation is only possible if
\begin{equation}
      \sigma < \frac{2}{\sqrt{3\kappa}}\left(\sqrt{|U_{\rm tv}|}
       -\sqrt{|U_{\rm fv}|} \right) \, .
\label{ineq}
\end{equation}
Here $\sigma$ is the surface tension in the bubble wall, $U_{\rm fv}$ and
$U_{\rm tv}$ are the energy densities of the false and true vacua, and 
$\kappa=8\pi G_N$.

As in the non-gravitational case, the CDL thin-wall approximation
requires that the two vacua be close in energy.  In addition, one must
require that the radius of curvature of the wall can be treated as
being constant as one moves through the bubble wall.
Recently it was shown~\cite{Masoumi:2016pqb} that there are regions of
parameter space that allow a new type of thin-wall regime in which the
latter requirement is violated.  In this case not only does the CDL
derivation of Eq.~(\ref{ineq}) fail, but also its very formulation
becomes ambiguous, because the surface tension is not well-defined.
In this paper we will show how this inequality can be generalized to
this new thin-wall regime.  Furthermore, we show how these bounds for
the thin-wall cases can be seen as special cases of a more general
bound, applicable even to bounce solutions that are in no sense
thin-wall.

We also discuss the case where the parameters of the theory are taken
to the boundary beyond which nucleation is quenched.  As the boundary
is approached, the bubble radius at nucleation increases without
bound.  When the critical values of the parameters are actually
achieved, the bounce solution is absent.  In its stead there is a
static planar domain wall~\cite{Vilenkin:1981zs,Ipser:1983db}.  Such
walls have been constructed as BPS solutions in supergravity
theories~\cite{Cvetic:1992bf,Cvetic:1992st,
  Cvetic:1996vr,Cvetic:1993xe,Ceresole:2006iq}, but they can also
arise as solutions that only possess what has been termed ``fake
supersymmetry''~\cite{Freedman:2003ax,DeWolfe:1999cp,Skenderis:2006jq}.
We will describe how this happens in our approach.  We will also
recall the related work of Abbott and
Park~\cite{Abbott:1985qa,Park:1986xa} connecting the existence of
bounces to the vacuum stability results of
Boucher~\cite{Boucher:1984yx}

The remainder of this manuscript is organized as follows.  In
Sec.~\ref{CDL} we review the CDL formalism, including their thin-wall
approximation.  In Sec.~\ref{beyond} the new thin-wall regime is
described and the generalization of Eq.~(\ref{ineq}) to this new
regime is derived.  In Sec.~\ref{AllBound} we derive the more general
bound that applies to all bounces.  In Sec.~\ref{critical} we discuss
the approach to the critical quenching limit where the Euclidean
bounce disappears and a static planar domain wall appears, and make
connections to supersymmetry. Section~\ref{conclude} summarizes our
results and comments on the extension to theories with multiple scalar
fields.  There is an appendix that addresses some special issues that 
arise when the false vacuum is Minkowskian.

\section{The CDL formalism}
\label{CDL}

We consider a theory with a real scalar field $\phi$ governed by a
potential $U(\phi)$ that has two metastable vacua at $\phi_{\rm tv}$
and $\phi_{\rm fv}$. The values of the potential at these vacua
satisfy $U_{\rm tv} < U_{\rm fv} \le 0$.
Thus, the higher false vacuum can be either Minkowski
or AdS, while the true vacuum is AdS.\footnote{Note that an AdS vacuum 
can correspond to a local maximum of $U(\phi)$, provided that the
Breitenlohner-Freedman bound is respected.}  The AdS vacua have 
characteristic lengths given by 
\begin{equation}
   \ell_{\rm fv} = \left(\frac{\kappa}{3} |U_{\rm fv}| \right)^{-1/2}
                = \left(-\frac{\kappa}{3} U_{\rm fv} \right)^{-1/2} 
\label{ellDef}
\end{equation}
and similarly for $\ell_{\rm tv}$.
Following CDL, we seek bounce solutions of the
Euclidean field equations.  Making the standard assumption of O(4)
symmetry, we can write the Euclidean metric in the form
\begin{equation}
     ds^2 = d\xi^2 + \rho(\xi)^2 \, d\Omega_3^2  \, .
\end{equation} 
For the cases we are considering, decays from a Minkowski or AdS
vacuum, $\rho(\xi)$ has a single zero and $\xi$ runs from 0 to $\infty$.
The bounce thus has $R^4$ topology, in contrast with the de Sitter 
bounces that are topologically four-spheres.

The Euclidean action can then be written in the form\footnote{The
  Gibbons-Hawking boundary term~\cite{Gibbons:1976ue}~ does not appear
  here because it is exactly canceled by the surface term from the
  integration by parts that removes the $\rho''$ that appears in the
  curvature scalar $R$.  In fact, the tunneling rate is unaffected by
  the inclusion or omission of the boundary term, because its
  contributions to the bounce action and the false vacuum action are
  equal, and so cancel in the tunneling exponent $B$~\cite{Masoumi:2016pqb}.}
\begin{equation}
    S= 2\pi^2 \int_0^\infty d\xi \, \left\{ \rho^3
    \left[\frac12\phi'^2 +U(\phi)\right] -\frac{3}{\kappa} 
    \left(\rho \rho'^2 +\rho\right) \right\}
\label{action}
\end{equation}
and a bounce must satisfy
\begin{equation}
   \phi'' + \frac{3\rho'}{\rho}\, \phi' = \frac{dU}{d\phi} \, ,
\label{phieq}
\end{equation}
\begin{equation}
   \rho'^2 = 1 +\frac{\kappa}{3}\rho^2
       \left[\frac12\phi'^2-U(\phi)\right]  \, ,
\label{rhoeq}
\end{equation}
subject to the boundary conditions
\begin{equation}
   \phi'(0) = 0 \, ,\qquad \phi(\infty)=\phi_{\rm fv} \, ,
    \qquad \rho(0)=0 \, ,
\end{equation}
where primes denote derivatives with respect to $\xi$.
Equations~(\ref{phieq}) and ~(\ref{rhoeq})
imply the useful equation
\begin{equation}
    \rho''= - \frac{\kappa}{3}\,\rho\left[\phi'^2 +U(\phi)\right] \, .
\label{ddrhoeq}
\end{equation}

We now note that $\rho(\xi)$ is a
monotonically increasing function.  To establish this, note first that
the boundary conditions imply that $\rho'(0)=1$. Requiring that the
bounce approach the pure false vacuum solution at large $\xi$ implies
that $\rho$ must asymptotically increase with $\xi$ either linearly
(for a Minkowski false vacuum) or exponentially (for an AdS false
vacuum).  If $\rho$ were not monotonic between these limits, it would
have a local minimum at some finite $\bar\xi$. This would require that
$\rho'(\bar\xi)=0$ and $\rho''(\bar\xi) > 0$. However, this cannot be,
since Eq.~(\ref{rhoeq}) shows that $U(\phi)$ must be positive at any
zero of $\rho'$, while Eq.~(\ref{ddrhoeq}) implies that $\rho''$ can
only be positive if $U(\phi)$ is negative.

We will find it useful to rewrite some of these results in terms of a
Euclidean pseudo-energy
\begin{equation}
    E = \frac12\phi'^2 - U(\phi) \, .
\label{Edef}
\end{equation}
Because of the $\phi'$ ``friction'' term in Eq.~(\ref{phieq}), this is not
conserved, but instead obeys
\begin{eqnarray}
    E' &=& - \frac{3 \rho'}{\rho}\, \phi'^2    \cr
    &=& - \phi'^2\sqrt{\frac{9}{\rho^2} + 3\kappa E }
\label{Eprime}
\end{eqnarray}
with the second line following with aid of Eq.~(\ref{rhoeq}).
We just showed that $\rho$ is a monotonically
increasing function of $\xi$.  It then follows that $E$ is
monotonically decreasing from an initial maximum 
$E(0) < |U_{\rm tv}|$ to an asymptotic minimum $E(\infty) = 
|U_{\rm fv}|$.  

The tunneling exponent $B$ is obtained by subtracting the action
of the homogeneous false vacuum solution from that of the bounce.
For configurations that satisfy Eq.~(\ref{rhoeq}) the
action can be rewritten as
\begin{eqnarray}
    S &=& 4\pi^2 \int_0^\infty d\xi\, \left[\rho^3 U(\phi) 
           - \frac{3}{\kappa}\rho \right]   \cr
      &=& 4\pi^2 \int_0^\infty d\rho \, \frac{1}{\rho'} \left[\rho^3 U(\phi)
           - \frac{3}{\kappa}\rho \right]  \, .  
\end{eqnarray}
The actions of the bounce and the false vacuum are both
divergent, so we must regulate the integrals.
Thus, we define
\begin{equation}
   S(L) = 4\pi^2 \int_0^L d\rho\, \frac{1}{\rho'} \left[
         \rho^3U(\phi) - \frac{3}{\kappa} \, \rho\right]
\end{equation}
and obtain a finite value for
\begin{equation}
    B = \lim_{L\to\infty} \left[S_{\rm bounce}(L) 
          - S_{\rm fv}(L) \right] \, .
\label{regularizedB}
\end{equation}

In particular, for an AdS false vacuum, 
with $\phi=\phi_{\rm fv}$
everywhere, 
\begin{equation}
     \rho_{\rm fv} = \ell_{\rm fv} \sinh(\xi/\ell_{\rm fv}) \, .
\label{rhoFV}
\end{equation}
Integrating the action density gives 
\begin{equation}
    S_{\rm fv}(L) = A_{\rm fv}(0,L)
\end{equation}
where 
\begin{eqnarray}
    A_{\rm fv}(\rho_1,\rho_2) &=& 4\pi^2 \int_{\rho_1}^{\rho_2} d\rho\, \frac{1}{\rho'}
    \left[ \rho^3U_{\rm fv} - \frac{3}{\kappa} \, \rho\right]  \cr
       &=& -\frac{4\pi^2}{\kappa}  \ell_{\rm fv}^2 \left[
     \left(1+ \frac{\rho_2^2}{\ell_{\rm fv}^2}\right)^{3/2} 
      -\left(1+ \frac{\rho_1^2}{\ell_{\rm fv}^2}\right)^{3/2}      
         \right] \, .
\end{eqnarray}

\subsection{The CDL thin-wall approximation}

In the CDL thin-wall approximation the bounce solution is divided into
three parts: an exterior region of pure false vacuum, an interior
region of pure true vacuum, and a thin wall that separates the two.
For such a configuration we can write
\begin{equation}
     B(\rho) =  B_{\rm exterior}(\rho) 
              +B_{\rm interior}(\rho)+B_{\rm wall}(\rho) 
\end{equation}
with $\rho$ being the curvature radius of the wall.  In the false
vacuum exterior region the actions of the bounce and the false vacuum
cancel completely, and so $B_{\rm exterior} =0$.  The contribution in
the interior region is the difference of true- and false-vacuum terms,
\begin{equation}
    B_{\rm interior} = A_{\rm tv}(0,\rho) - A_{\rm fv}(0,\rho) \, .
\end{equation}
Finally, we have the contribution from the wall, which can be written in 
the form
\begin{equation}
  B_{\rm wall} = 2\pi^2 \rho^3 \sigma
\label{CDLwall}
\end{equation}
where the surface tension $\sigma$ is given by the flat-spacetime
expression\footnote{The absence of gravitational corrections in the
  CDL expression for the surface tension will be justified in
  Sec.~\ref{beyond}.}
\begin{equation}
       \sigma = 2 \int_{\rm wall} d\xi 
        \left[U(\phi_{\rm bounce}) - U_{\rm fv} \right]
\label{sigmaDef}
\end{equation}
and the integration over $\xi$ is restricted to the wall 
region\footnote{More precisely, CDL replace $U$ in the integral
by a function $U_0$ that has minima at $\phi_{\rm tv}$ and $\phi_{\rm fv}$
and is equal to $U_{\rm fv}$ at both minima.  In the thin-wall limit
the effect of this replacement is higher order.}

It is crucial here that the field profile in the wall, and hence
$\sigma$, are to a good approximation independent of $\rho$, a
consequence of the fact that the bounce radius is much greater than
the thickness of the wall.  This approximation becomes better as the
difference between the true and false vacuum energies decreases, not
because the wall gets thinner (it doesn't), but because the bounce
gets bigger.  Indeed, one might term this the ``large-bounce
approximation''.

Note that Eq.~(\ref{CDLwall}) implicitly assumes that $\rho$ is
essentially constant as one moves through the wall.  The CDL thin-wall
analysis is only valid if this is the case.  In Sec.~\ref{beyond} we
will consider thin-wall configurations for which this assumption
fails.

The bounce is obtained by requiring that the wall radius be a stationary
point $\bar\rho$ of $B(\rho)$. Setting $dB/d\rho = 0$ leads to 
\begin{eqnarray}
  \sigma &=& \frac{2}{\kappa} 
    \left(\sqrt{ \frac{1}{\ell^2_{\rm tv}} - \frac{1}{\bar\rho^2} }
  - \sqrt{ \frac{1}{\ell^2_{\rm fv}} - \frac{1}{\bar\rho^2} } \right)\cr
    &<&  \frac{2}{\kappa}  \left( \frac{1}{\ell_{\rm tv}}
       -\frac{1}{\ell_{\rm fv}} \right)  \, .
\label{ineq2}
\end{eqnarray}
with the bound being approached in the limit $\bar\rho \to \infty$.
Using Eq.~(\ref{ellDef}) to rewrite the inequality on the second line
yields the bound in Eq.~(\ref{ineq}).

\section{Thin-wall bounces beyond CDL}
\label{beyond}

Reference~\cite{Masoumi:2016pqb} examined tunneling in the more general case where
the true and false vacua are not close in energy and the conditions
for CDL's thin-wall approximation are not met. It was found that as the
mass scales in the potential are increased, making gravitational
effects stronger, a new type of thin-wall regime emerges. More
specifically, for any given potential one can define a quantity $\beta$ as
the ratio of a mass scale in the potential to the Planck
mass.  Gravitational effects become stronger as $\beta$ is increased.
Eventually, as $\beta$ approaches a critical value, the bounce 
radius tends to infinity.  In the limit the bounce solution
disappears, tunneling is completely quenched, and the false vacuum is
stabilized.

In this new thin-wall regime the scalar field profiles are
qualitatively similar to those in the CDL thin-wall bounces.  There is
an interior region, $0<\xi<\xi_1$, that is approximately pure true
vacuum, an exterior region, $\xi_2 < \xi < \infty$, that is almost pure
false vacuum, and a narrow transition region, or wall, that separates
the two, with the wall thickness\footnote{This definition of the wall
thickness differs from that used in~\cite{Masoumi:2016pqb}, which only
  included regions where $U(\phi) > U_{\rm fv}$.} $\Delta\xi = \xi_2
-\xi_1$ being small compared to $\xi_1$.  However, they differ from
their CDL counterparts in that $\rho(\xi)$ grows considerably as one
passes through the wall, and so cannot be approximated as being
constant.

As with the CDL thin-wall approximation, it is convenient to write the
tunneling exponent as the sum of three terms, each of which is the
difference between a bounce action term and a corresponding false
vacuum term.  In order to be consistent
with the form of the long-distance regulator of the action integrals,
Eq.~(\ref{regularizedB}), the corresponding regions of the bounce and the false
vacuum must be defined by values of $\rho$, rather than $\xi$.
Thus, if the wall in the bounce solution runs between $\xi_1$ and $\xi_2$,
then the corresponding false vacuum region is bounded by 
$\rho_1 \equiv \rho(\xi_1)$ and $\rho_2 \equiv \rho(\xi_2)$.  

With this understanding, we obtain for the interior region, $\rho<\rho_1$,
\begin{eqnarray}
    B_{\rm interior} &=&  A_{\rm tv}(0,\rho_1) - A_{\rm fv}(0,\rho_1)
     \cr   &=& - \frac{4\pi^2}{\kappa} \left\{ \ell_{\rm tv}^2
     \left[\left(1+ \frac{\rho_1^2}{\ell_{\rm tv}^2}\right)^{3/2}
       -1\right]  
       -\ell_{\rm fv}^2
     \left[\left(1+ \frac{\rho_1^2}{\ell_{\rm fv}^2}\right)^{3/2}
       -1\right]  \right\} \, ;
\end{eqnarray}
i.e., the CDL result with $\rho =\rho_1$.    
In the exterior region the actions of the bounce and the false vacuum 
exactly cancel, so $B_{\rm exterior} = 0$.

In the wall region we have
\begin{eqnarray}
   B_{\rm wall} &=& 4\pi^2\int_{\rho_1}^{\rho_2}d\rho \left\{ \frac{1}{\rho_b'}
                 \left[\rho^3 U(\phi_b) - \frac{3}{\kappa} \rho \right]
       - \frac{1}{\rho_{\rm fv}'}
         \left[\rho^3 U_{\rm fv} - \frac{3}{\kappa} \rho \right] \right\}   \cr
         &=&
 4\pi^2\int_{\xi_1}^{\xi_2} d\xi \, 
          \left[\rho^3 U(\phi_b) - \frac{3}{\kappa} \rho \right]
        - A_{\rm fv}(\rho_1,\rho_2) \, .
\label{WallContrib}
\end{eqnarray}
In the first line the subscripts on $\rho'$ and in the potential term
indicate that in the first term these are to be evaluated from the 
bounce solution, and in the second term from the pure false 
vacuum solution.  

This expression for $B_{\rm wall}$ should reduce to the CDL result in
the limit where $\Delta \rho = \rho_2- \rho_1$ is small.  To verify
this, we write the false vacuum contribution to
Eq.~(\ref{WallContrib}) as
\begin{equation}
  - A_{\rm fv}(\rho_1,\rho_1+\Delta\rho) 
    = \frac{4 \pi^2}{\kappa} (3 \rho_1)\sqrt{1+ 
       \frac{\rho_1^2}{\ell_{\rm fv}^2}} \, \Delta\rho
     + O[(\Delta\rho)^2]  \, .
\end{equation}
Now $\Delta\rho = \rho'(\xi_1)\Delta \xi$. In the false 
vacuum, $\rho'= \sqrt{1 + \rho^2/\ell_{\rm fv}^2}$. Using these
facts, we obtain
\begin{eqnarray}
  - A_{\rm fv}(\rho_1,\rho_1+\Delta\rho) &=&
    \frac{4 \pi^2}{\kappa} (3 \rho_1) 
   \left(1+  \frac{\rho_1^2}{\ell_{\rm fv}^2} \right) \Delta \xi
     + O[(\Delta\xi)^2] \cr
    &=& 4 \pi^2 \left( \frac{3\rho_1}{\kappa} - \rho_1^3 U_{\rm fv}
     \right)\Delta \xi + O[(\Delta\xi)^2]  \, .
\end{eqnarray}
Combining this result with the contribution from the bounce, and
working to first order in $\Delta\xi$, we recover the CDL result, with
the surface tension given by Eq.~(\ref{sigmaDef}).  Note that this
justifies CDL's use of the flat-spacetime expression for the surface
tension.

Let us now return to the more general case, with $\rho_2-\rho_1$ not
assumed to be small.  We can no longer approximate $\rho$ as being
constant through the wall.  One consequence is that the identification
of a surface tension becomes problematic.  One usually defines surface
tension in terms of an energy per unit area (or action per unit
hypersurface area).  Because $\rho(\xi)$ grows in the wall, the area
of the outer surface of the wall is larger than that of the inner
surface of the wall.  Which, if either, should be used?  In fact, it
is not even obvious that the wall action can be written as the product
of an area and a radius-independent factor.

To answer these questions we need to examine the form of these new
thin-wall solutions in more detail.  The scalar field at the center of
the bounce, $\phi(0)$, is very close to $\phi_{\rm tv}$.  The
field remains close to $\phi_{\rm tv}$ until $\xi \approx \xi_1$, so for
the interior region, $\xi \lesssim \xi_1$, we have, analogously to
Eq.~(\ref{rhoFV}),
\begin{equation}                                                          
    \rho \approx \ell_{\rm tv} \sinh(\xi/\ell_{\rm tv}) \, .
\end{equation}  
If gravitational effects are made stronger by increasing $\beta$, 
$\xi_1$ increases and $\rho_1$ grows exponentially.

In the near critical regime the growth of $\rho$ in the interior
region is such that at $\xi_1$ the first term on the right-hand side
of Eq.~(\ref{rhoeq}) can be neglected.  If $\rho_1 \gg \ell_{\rm fv}$
this remains true throughout the wall, and beyond.  We can then write
\begin{equation}
    \frac{\rho'}{\rho}
      = \sqrt{\frac{\kappa}{3}} \sqrt{\frac12 \phi'^2 - U(\phi)}
\label{drhoOverRhoApprox}
\end{equation}
so that Eq.~(\ref{phieq}) becomes 
\begin{equation}
   \phi'' + \sqrt{3\kappa} \sqrt{\frac12 \phi'^2 - U(\phi)} \, \,
         \phi' = \frac{dU}{d\phi} \, .
\label{phiOnlyEq}
\end{equation}
Note that $\rho$ does not appear in this equation. Hence the profile
of $\phi(\xi)$ in the wall is independent of $\rho$.

Furthermore, integration of Eq.~(\ref{drhoOverRhoApprox}) gives
\begin{equation}
     \rho(\xi) = \rho_1 \,e^{G(\xi)}
\end{equation}
where
\begin{equation}
   G(\xi) =  \sqrt{\frac{\kappa}{3}} 
       \int_{\xi_1}^\xi d\xi\,\sqrt{\frac12 \phi'^2 - U(\phi)} 
\label{rhoExp}
\end{equation}
is also independent of $\rho$.
This allows us to rewrite Eq.~(\ref{WallContrib}) as
\begin{equation}
   B_{\rm wall} = 4\pi^2 \int_0^{\ln(\rho_2/\rho_1)} dG\, \left\{\frac{1}{G_b'}
          \left[\rho_1^3 U(\phi_b) \, e^{3G}
     - \frac{3}{\kappa} \rho_1 \, e^{G}\right]
     -\sqrt{\frac{3}{\kappa}}\frac{1}{\sqrt{-U_{\rm fv}}}\left[\rho_1^3 U_{\rm fv} e^{3G}
      - \frac{3}{\kappa} \rho_1 e^G \right] 
        \right\}  \, .
\end{equation}

In the limit of large bounce radius ($\rho_1 \gg l_{\rm fv}$), the
terms cubic in $\rho_1$ dominate.  Keeping only these, we have
\begin{equation}
   B_{\rm wall} = 4\pi^2 \rho_1^3\sqrt{\frac{3}{\kappa}}
     \int_0^{ \ln(\rho_2/\rho_1)} dG\, e^{3G} \left[ 
     \frac{U(\phi_b)}{\sqrt{\frac12 {\phi'}_b^2-U(\phi_b)}} 
       + \sqrt{- U_{\rm fv}} \right]  \, .
\end{equation}
This suggests that we write
\begin{equation}
   B_{\rm wall} = 2 \pi^2 \rho_1^3 \tilde \sigma
\end{equation}
where $2 \pi^2 \rho_1^3$ is the area of the inner surface of the wall
and
\begin{equation}
  \tilde\sigma = \sqrt{\frac{12}{\kappa}}
     \int_0^{\ln(\rho_2/\rho_1)} dG\, e^{3G} \left[
  \frac{U(\phi_b)}{\sqrt{\frac12 {\phi'}_b^2-U(\phi_b)}}
       + \sqrt{- U_{\rm fv}} \right]
\label{sigmaTilde}
\end{equation}
can be viewed as a generalization of the CDL surface tension 
$\sigma$.\footnote{In the CDL expression for the surface tension,
  Eq.~(\ref{sigmaDef}), the integrand is everywhere positive so
  $\sigma$ is manifestly positive.  This is not the case for
  $\tilde\sigma$. Indeed, the integrand in Eq.~(\ref{sigmaTilde}) is
  negative in the lower part of the integration range and positive in
  the upper part.  In the next section we will show that this
  expression for $\tilde \sigma$ is a special case of a more general
  expression that is manifestly positive.}
(Note that, like $\sigma$, it is independent of $\rho$.)  With this
definition, the total expression for $B$ takes the same form as in the
CDL thin-wall limit, but with the replacements $\bar\rho \to \rho_1$
and $\sigma \to \tilde\sigma$. The line of reasoning that led to
Eq.~(\ref{ineq2}) and then to Eq.~(\ref{ineq}) now leads
to
\begin{equation}
      \tilde\sigma < \frac{2}{\sqrt{3\kappa}}\left(\sqrt{|U_{\rm tv}|}
       -\sqrt{|U_{\rm fv}|} \right) \, .
\label{ineq3}
\end{equation}

\section{A bound for all bounces}
\label{AllBound}

We have obtained upper bounds on the surface tension for both the
thin-wall approximation of CDL and the generalized thin-wall regime 
of~\cite{Masoumi:2016pqb}.  However, thin-wall bounces of either type are 
special cases.  There are bounce solutions that are not in any sense
thin-wall, including some for which it is difficult to even define a
surface tension.  This raises the question of whether there is a more
general bound that applies to all bounces and that reduces to
Eqs.~(\ref{ineq}) and (\ref{ineq3}) in the appropriate limits.

We now show that there is.  To begin, we recall the 
definition of the pseudo-energy, Eq.~(\ref{Edef}), and the expression
for its derivative, Eq.~(\ref{Eprime}).  If follows from the latter
that
\begin{equation}
       \frac{d\sqrt{E}}{d\xi} = - \frac{\sqrt{3\kappa}}{2} 
         \sqrt{1 + \frac{3}{\kappa E \rho^2} }  \, \, \phi'^2 \, .
\label{sqrtderiv}
\end{equation}
Integrating this, we find that 
\begin{eqnarray}
    \sqrt{E(0)} - \sqrt{E(\infty)} &=& \frac{\sqrt{3\kappa}}{2}
    \int_0^\infty d\xi \,\sqrt{1 + \frac{3}{\kappa E \rho^2} } 
        \,\, \phi'^2   \cr 
    &>& \frac{\sqrt{3\kappa}}{2} \int_0^\infty d\xi \, \phi'^2 \, .
\end{eqnarray}
Noting that $E(0) \le |U_{\rm tv}|$ and recalling that 
$E(\infty) = |U_{\rm fv}|$, we have
\begin{equation}
        \int_0^\infty d\xi \, \phi'^2_b < \frac{2}{\sqrt{3\kappa}}
       \left( \sqrt{|U_{\rm tv}|} -\sqrt{|U_{\rm fv}|} \right)     \, .  
\label{generalineq}
\end{equation}

This inequality is exact, and does not depend on any approximations.
It therefore applies to any bounce solution.  In particular, it should
reduce to our previous results for thin-wall bounces.  In these
bounces $\phi'$ is taken to vanish outside the wall region, so the
integration can be restricted to the range $\xi_1<\xi<\xi_2$.  In the
CDL thin-wall approximation the bounce profile in the wall region is
approximately that of a (1+1)-dimensional kink.
Equation~(\ref{sigmaDef}) gives the surface tension in terms of an
integral of the potential $U(\phi_b)$. A virial
theorem~\cite{Weinberg:2012pjx} relates this to the integral of
$\phi'^2$ and shows that the bounds of Eqs.~(\ref{ineq}) and
(\ref{generalineq}) are equivalent within the accuracy of the
approximation.

For the new thin-wall case, demonstrating the equivalence of
Eqs.~(\ref{ineq}) and (\ref{ineq3}) requires a bit more work.
We begin by noting the identity
\begin{equation}
   \frac{1}{3\kappa} \int_{\xi_1}^{\xi_2} d\xi \,
        \frac{d}{d\xi}\left(\rho^3 \sqrt{E} \right) 
      = \int_{\xi_1}^{\xi_2} d\xi \, \rho^3 U(\phi_b) \sqrt{1 + 
              \frac{3}{\kappa E \rho^2} }  \, ,
\label{integral1}
\end{equation}
which is obtained by evaluating the derivative inside the integral
on the left-hand side and using Eq.~(\ref{sqrtderiv}).

Alternatively, using the fact that the integrand is a total
derivative gives
\begin{eqnarray}
   \frac{1}{3\kappa} \int_{\xi_1}^{\xi_2} d\xi \,
        \frac{d}{d\xi}\left(\rho^3 \sqrt{E} \right) 
    &=& \frac{1}{3\kappa} \left[ -(\rho_2^3 - \rho_1^3)\sqrt{E_2}
      - \rho_1^3 (\sqrt{E_2}- \sqrt{E_1} ) \right]   \cr
    &=&  \frac{1}{3\kappa} \left[- (\rho_2^3 - \rho_1^3)\sqrt{E_2}
    -\rho_1^3  \int_{\xi_1}^{\xi_2} d\xi \, \frac {d\sqrt{E}}{d\xi} \right] \cr
   &=& - \frac{1}{\kappa\ell_{\rm fv}} (\rho_2^3 - \rho_1^3)
         -\frac12 \rho^3_1  \int_{\xi_1}^{\xi_2} d\xi \,
         \phi'^2 \sqrt{1 + \frac{3}{\kappa E\rho^2}}  \, .
\label{integral2}
\end{eqnarray}
In the last equality we have used the definition of the AdS length,
Eq.~(\ref{ellDef}), and the fact that $E_2 = -U_{\rm fv}$.
Comparing Eqs.~(\ref{integral1}) and (\ref{integral2}), we have 
\begin{equation}
    \int_{\xi_1}^{\xi_2} d\xi \, \rho^3 U(\phi_b) \sqrt{1 +
      \frac{3}{\kappa E\rho^2}} 
     + \frac{1}{\kappa\ell_{\rm fv}} (\rho_2^3 - \rho_1^3)
    = -\frac12 \rho^3_1  \int_{\xi_1}^{\xi_2} d\xi \,
     \phi'^2  \sqrt{1 + \frac{3}{\kappa E\rho^2}} \, .
\label{IntEqualInt}
\end{equation}

In the range of integration, $\xi_1 \le \xi \le \xi_2$, the pseudo-energy
satisfies $E \ge E_2 = |U_{\rm fv}|$, while $\rho \ge \rho_1$. It follows
that
\begin{equation}
  \frac{3}{\kappa E\rho^2}  < \frac{\ell_{\rm fv}^2}{\rho_1^2} \, . 
\end{equation}
Hence in the limit of large bounce radius, $\rho_1 \gg                          
\ell_{\rm fv}$, 
the square roots in Eq.~(\ref{IntEqualInt}) can be set equal to
unity, giving
\begin{equation}
       \int_{\xi_1}^{\xi_2} d\xi \, \rho^3 U(\phi_b) 
     + \frac{1}{\kappa\ell_{\rm fv}} (\rho_2^3 - \rho_1^3)
      = -\frac12 \rho^3_1  \int_{\xi_1}^{\xi_2} d\xi \, \phi'^2 \, .
\label{intEqualint}
\end{equation}
In this same limit
Eq.~(\ref{WallContrib}) reduces to
\begin{eqnarray}
   B_{\rm wall} &=& 4\pi^2 \int_{\xi_1}^{\xi_2} d\xi \,\rho^3 U(\phi_b)
    + \frac{4\pi^2}{\kappa \ell_{\rm fv}} (\rho_2^3 - \rho_1^3)   \cr
     &=& 2 \pi^2 \rho^3_1  \int_{\xi_1}^{\xi_2} d\xi \, \phi'^2   
\end{eqnarray}
where the second line follows from Eq.~(\ref{intEqualint}).
Dividing by the surface area, $2 \pi^2 \rho^3_1$, gives
\begin{equation}
   \tilde\sigma =  \int_{\xi_1}^{\xi_2} d\xi \, \phi'^2   
\label{sigEqual}
\end{equation}
and demonstrates the equivalence of Eqs.~(\ref{ineq3}) and (\ref{generalineq})
in this regime.

\section{Bounces and walls in the critical limit}
\label{critical}

It is instructive to examine the behavior of the bounce solution as
the parameters of the theory approach the critical limit where the
nucleation of AdS bubbles is totally quenched.  As this limit
is approached the wall radius of the bounce solution grows
without bound, with both $\xi_1$ and $\rho_1$ diverging.  When the 
parameters are actually at their limiting values, there is no 
Euclidean bounce at all.  

Now let us focus on the fixed time slice through the center of the
bounce, $t=0$, when the bubble is nucleated.  The bubble is
instantaneously at rest, while the spatial part of the metric is
\begin{equation}
      d\ell^2 = d\xi^2 +\rho(\xi)^2 d\Omega^2_2
\end{equation}
with $\rho(\xi)$ taken over from the bounce and $\xi$ 
again being a radial coordinate.  The limit of infinite 
bubble radius can be viewed, in a certain sense, as a 
planar wall separating two metastable vacua.

The metric for a static planar wall can be written as
\begin{equation}
    ds^2 = A(z)(-dt^2 +dx^2 +dy^2) +dz^2
\end{equation}
with $z$ being the spatial coordinate orthogonal to the wall.  For our
theory with a single scalar field, the field equations are
\begin{equation}
    A'^2 = \frac{\kappa}{3} A^2 \left(\frac12 \phi'^2 - U  \right)
\end{equation}
\begin{equation}
    \phi'' + \frac{3A'}{A} = \frac{dU}{d\phi}
\end{equation}
with primes indicating differentiation with respect to $z$.  These 
are to be solved subject to the boundary conditions that $\phi$ 
take its false (true) vacuum value at $z=\infty$ ($z=-\infty$).

If we make the correspondence 
\begin{equation}
    \xi \longleftrightarrow z \, ,  \qquad \rho \longleftrightarrow A  \, ,
\end{equation}
these equations differ from Eqs.~(\ref{phieq}) and (\ref{rhoeq})
only by the omission of the factor of unity on the right-hand
side of Eq.~(\ref{rhoeq}).  The boundary conditions at $\xi=\infty$
and $z=\infty$ agree.  Although those at $\xi=0$ and $z=-\infty$
differ, they coincide in the limit of infinite bubble radius.

There is an interesting connection with supersymmetry when the parameters
are near the critical limit.
Let us suppose that we are given $\phi(\xi)$ and $\rho(\xi)$
satisfying the bounce equations Eqs.~(\ref{phieq}) and (\ref{rhoeq}).
Because $\phi$ is a monotonic function of $\xi$, we can view $\xi$,
and therefore $\phi'$, $E$, and $\rho$, as functions of $\phi$ in the
neighborhood of the bounce solution.  We can then define a function
$f(\phi)$ by
\begin{equation}
    f(\phi)^2 = \frac{1}{3\kappa} \left( \frac12 \phi'^2 -U \right)
   = \frac{1}{3\kappa}E  \, .
\label{defOFf}
\end{equation}
The derivative of $f$ with respect to $\phi$ is
\begin{eqnarray} 
    \frac{df}{d\phi} &=& \frac{df/d\xi}{d\phi/d\xi}   \cr
         &=& \frac{1}{\sqrt{3\kappa}} \frac{d\sqrt {E}}{d\xi}\frac{1}{\phi'}
 \cr &=& -\frac12 \phi' \alpha^{-1}
\label{dfdphi}
\end{eqnarray}
where the last line follows from Eq.~(\ref{sqrtderiv})
and
\begin{equation}
   \alpha(\phi) = \left(1 + \frac{1}{\kappa^2 \rho^2 f^2} \right)^{-1/2} \, .
\end{equation}

Solving Eq.~(\ref{dfdphi}) for $\phi'$ and substituting the result into
Eq.~(\ref{defOFf}) leads to 
\begin{equation}
    U = 2\alpha^2\left(\frac{df}{d\phi}\right)^2
           - 3 \kappa f^2  \, .
\label{quasiSugra}
\end{equation}
If $\alpha$ were equal to unity, this would be the form for the
potential in a supergravity theory, with $f$ being the
superpotential. In fact, $\alpha \approx 1$ wherever $\rho \gg
\ell_{\rm fv}$.  Thus, for a near-critical bounce the deviation from
the supersymmetric form is confined to a region of approximate true
vacuum in the center of the bounce.\footnote{This assumes that the
  false vacuum is AdS.  The case of a Minkowski false vacuum is
  addressed in the appendix.}  

Repeating the calculation for the static planar wall, one finds that
$\alpha = 1$ everywhere in space.  The domain wall would then have the form 
of a supersymmetric wall interpolating
between isolated vacua of an ${\cal N}=1$ supergravity
potential, with the disappearance of the bounce solution guaranteeing the
stability of each of these vacua against decay by nucleation of
bubbles of the other.
Note, however, that our construction is restricted to the
interval of field space between the two vacua; the form of $U(\phi)$
outside this interval is unconstrained and need not be derivable from 
a superpotential.

An alternative path to demonstrating vacuum stability is to prove a positive
energy theorem or a BPS bound. In the presence of gravity, Boucher
\cite{Boucher:1984yx}, generalizing Witten's work
\cite{Witten:1981mf,Nester:1982tr}, gave the following criteria: an extremum
$\bar{\phi}$ of a potential $U(\phi)$ with $U(\bar{\phi}) < 0$ is
stable if there exists a real function $W(\phi)$ such that
\begin{equation}
     2 \left(\frac{d W}{d \phi}\right)^2- 3 \kappa W^2
     = U(\phi) \hspace{4ex} \forall \phi 
\label{boucher}
\end{equation}
and
\begin{equation}
    W(\bar{\phi})=\left[ -U(\bar{\phi})/{3 \kappa}\right]^{1/2} \, .
\end{equation}
These criteria make no direct reference to bubble nucleation.  That
connection was drawn by Abbott and Park~\cite{Park:1986xa}. Given a
potential $U(\phi)$, one can obtain $W(\phi)$ by integrating
Eq.~(\ref{boucher}), provided that $U+3\kappa W^2$ remains positive.
Abbott and Park showed that if the latter becomes negative at some
$\phi=\phi_s$ before one reaches an extremum of $U$, then $\phi_s$ is
the starting point $\phi(0)$ for a bounce solution that governs the
decay via bubble nucleation of the $\bar\phi$ vacuum.

\section{Concluding remarks}
\label{conclude}

Gravity can quench the nucleation of bubbles in a Minkowski or AdS
false vacuum.  This result was proven analytically by Coleman and De
Luccia in a thin-wall limit where the energy difference between the
true and false vacua is small.  Their analysis implies an inequality,
Eq.~(\ref{ineq}), that relates the surface tension $\sigma$ and the
true- and false-vacuum energies. It is essential for this inequality
that $\sigma$ is independent of the radius of curvature of the bounce 
wall, and that the contribution of the wall to the Euclidean action
is the product of $\sigma$ and the surface area of the wall.

Subsequently, numerical analyses have generalized this claim to a
wider variety of potentials \cite{Samuel:1991dy, Bousso:2006am,
  Aguirre:2006ap, Masoumi:2016pqb}. As the boundary beyond which
nucleation is quenched is approached, the bubble radius at nucleation
increases without bound and a new thin-wall regime emerges
\cite{Masoumi:2016pqb}. This new thin-wall regime differs from the CDL
thin-wall regime in that the wall radius of curvature $\rho$ grows
exponentially as one moves through the bubble wall. It is then far
from obvious that the wall action can be decomposed as the product of
a surface tension and an area. Indeed, it is not even clear how the
area should be defined.  In Sec.~\ref{beyond} we used the fact that
the matter field profile in the wall is independent of the curvature
of the wall to define a modified surface tension $\tilde \sigma$ that
is $\rho$-independent.  We were then able to show analytically that 
the wall action is $\tilde\sigma$ times the area of the inner surface
of the bounce wall.  Closely following the reasoning of CDL then led
to the bound of Eq.~(\ref{ineq3}) on $\tilde\sigma$.

In Sec.~\ref{AllBound} we proved that the upper bounds on the surface
tensions in the two thin-wall regimes are limiting cases of a more
general bound, Eq.~(\ref{generalineq}), that is satisfied by all
bounces. Gravity sources a frictional force that depletes the
pseudo-energy as it evolves through the radial direction of the
bounce. This bound expresses the constraint on this frictional force
that is required if the bounce is to interpolate between the two vacua.

The work presented here was limited to single-field potentials, but the
proof of the bound of Eq.~(\ref{generalineq}), the existence of the
generalized thin-wall regime~\cite{Masoumi:2016pqb} and the
definition of surface tension in this regime, Eq.~(\ref{sigmaTilde}), should
extend to multifield potentials.  With $N$ scalar fields $\phi_i$ the bounce
should satisfy
\begin{eqnarray}
{\phi_i}'' + \frac{ 3 \rho'}{\rho} \phi_i' = \frac{d U}{d \phi_i} \, , \\
\rho'^2 = 1 + \frac{\kappa}{3}\rho^2 
  \left[  \sum_i \frac{1}{2} \phi_i'^2 - U (\phi_i) \right] \, .
\end{eqnarray} 
Solving these will lead to a trajectory through field space of the
form $\phi_i(\xi) = g_i(\xi)$.  In the neighborhood of this bounce
trajectory we can introduce a coordinate $\Phi$ along the trajectory,
defined by 
\begin{equation}
   d \Phi^2 = \sum_{i=1}^{N} dg_i^2    
\end{equation}
together with $N-1$ fields normal to the trajectory. 

In these new coordinates the action along the bounce 
\begin{equation}  S= 2 \pi^2 \int_0^{\infty} d\xi \, \left\{ \rho^3 \left[
  \frac{1}{2} \Phi'^2 - U(\Phi) \right]-\frac{3}{\kappa} ( \rho
\rho'^2+\rho)\right\} 
\end{equation}
only depends on $\Phi$.  The form of $U(\Phi)$ depends on the explicit
bounce solution, but the only information that was assumed about the
potential was that it was a continuous function interpolating between
$U_{\textrm{tv}}$ and $U_{\textrm{fv}}$.  The formal arguments made in
Secs.~\ref{beyond} and \ref{AllBound} should extend to the field
$\Phi$, and the bound of Eq.~(\ref{generalineq}) will be satisfied.

In the single-field case $U(\phi)$ approached a supersymmetric form as
the parameters of the theory approached the boundary where nucleation
was quenched.  Precisely on this boundary, where the false vacuum
becomes stable against decay by tunneling, a static planar domain wall
appears.  $U(\phi)$ takes on the supergravity form and can be written
in terms of a fake superpotential $W(\phi)$.  However, this form is only
guaranteed on the interval in field space lying between the true and
false vacua.  For the multifield case similar arguments allow one
to define a fake superpotential, but only along the trajectory of
the bounce in field space and only encoding a dependence on $\Phi$, but
not on the fields normal to the trajectory.

Even though these restrictions weaken the connection between stability
and supersymmetry, recent
claims~\cite{Ooguri:2016pdq,Freivogel:2016qwc,Banks:2016xpo} based on
the weak gravity conjecture~\cite{ArkaniHamed:2006dz} assert the
instability of all non-supersymmetric vacua in a UV complete
theory. Said differently, theories for which stable domain walls do
not obey Eq.~(1.1) live in the swampland, with gravity never strong
enough to quench a decay.  See \cite{Danielsson:2016rmq,
  Ooguri:2017njy, Danielsson:2016mtx,Danielsson:2017max} for more
examples of this instability.

\begin{acknowledgments}

We thank Adam Brown and Dan Freedman for useful discussions.
S.P. thanks the members of the Stanford Institute of Theoretical Physics for their
hospitality. E.W. thanks the Korea Institute for Advanced Study and
the Department of Applied Mathematics and Theoretical Physics of the
University of Cambridge for their hospitality.  Part of this work was
done at Aspen Center for Physics, which is supported by U.S. National
Science Foundation grant PHY-1607611. This work was also supported by
the U.S. National Science Foundation under Grants No.PHY-1518742, PHY-1521186 and
PHY-1620610 and by the U.S. Department of Energy under Grant
No. DE-SC0011941.

\end{acknowledgments}

\appendix*

\section{Minkowski false vacuum}

At some points in our analysis we have relied on the condition that
$\rho_1 \gg \ell_{\rm fv}$.  With an AdS false vacuum this 
can always be achieved by going sufficiently close to the critical
limit.  This is not the case if the false vacuum is Minkowskian, with
$U_{\rm fv} =0$ implying $\ell_{\rm fv} = \infty$.  Here we examine
the consequences of this fact. 

In our analysis we used this inequality to argue that in the wall and
in the entire exterior region the right-hand side of Eq.~(\ref{rhoeq})
was dominated by the second term.  This led to the conclusion that the
field profile in the wall was independent of $\rho$.  It also allowed
us to approximate the square roots in Eq.~(\ref{IntEqualInt}) as
unity, which led to Eq.~(\ref{sigEqual}) and the equivalence of
Eqs.~(\ref{ineq3}) and (\ref{generalineq}).

Now suppose that the false vacuum is Minkowskian.  By going
sufficiently close to the critical limit we can always ensure that the
second term dominates the right-hand side of Eq.~(\ref{rhoeq}) near
the inner edge of the wall, but as $E$ decreases with
increasing $\xi$ there comes a $\bar\xi$ where
\begin{equation}
    \frac{\kappa}{3} E = \frac{\kappa}{3} \left[ \frac12 \phi'^2 -
      U\right] =\frac{1}{\rho^2}  \, .
\end{equation}
For a thin-wall bubble (of either kind) $\rho$ is already
exponentially large at $\xi_1$, which means that $E$, $U(\phi_b)$, and
$\phi'^2$ will all have become exponentially small by the time that
$\bar\xi$ is reached.  Thus, the nontrivial part of the field profile
will lie in the region $\xi < \bar\xi$ and will be independent of
$\rho$, just as with an AdS false vacuum.  Near the critical limit
$\bar\xi$ will lie outside the wall (i.e., $\bar\xi > \xi_2$), so the
square roots in Eq.~(\ref{IntEqualInt}) will remain close to unity for
the entire range of integration.

We saw in Sec.~\ref{critical} that the potential closely approximates
the supergravity form wherever $\alpha\approx 1$ or, equivalently,
$\kappa E \rho^2 \gg 1$.  With an AdS false vacuum $E$ has a nonzero
lower bound, and so for near-critical bounces these conditions hold
everywhere except for a region of approximate true vacuum in the
center of the bounce.  With a Minkowski false vacuum $E$ tends to zero
as $\rho\to\infty$, so the large-distance behavior requires closer
examination.  To begin, note that 
\begin{eqnarray}
    (E \rho^2)' &=& 2\rho \rho' E +\rho^2 E'  \cr
        &=& \rho \rho' \left(2E - 3\phi'^2 \right)  \cr
        &=& -2 \rho \rho' \left(U+\phi'^2 \right)  \, .
\end{eqnarray}
When $\phi$ is close to its value at the Minkowski false vacuum
minimum, $U(\phi)$ is positive, so the quantity in parentheses on the
last line is positive definite, but exponentially decreasing in
magnitude. Near the false vacuum $\rho$ grows linearly with $\xi$, so
the integrated decrease in $E \rho^2$ as $\xi\to\infty$ is finite.
Hence, for near-critical bounces where $E \rho^2$ is large even near the 
end of the wall region, it will remain large as $\xi$ increases.  Just
as with an AdS false vacuum, in a near-critical bounce $\alpha$ will 
only deviate from unity near the center of the bounce, far from the 
wall.

\end{document}